%% file: newneu.tex
\documentclass[12pt]{article}

\usepackage{amsmath}
\usepackage{times}
\usepackage{xspace}
\usepackage{array}
\usepackage{epsfig}
\usepackage{latexsym}
\usepackage{amssymb}
\newcommand{\numub}{\ensuremath{\overline{\nu}_\mu}\xspace}
\newcommand{\nueb}{\ensuremath{\overline{\nu}_e}\xspace}
\newcommand{\numu}{\ensuremath{\nu_\mu}\xspace}
\newcommand{\nue}{\ensuremath{\nu_e}\xspace}

\newcommand{\nutau}{\ensuremath{\nu_\tau}\xspace}

\newcommand{\nubar}{\ensuremath{\overline{\nu}}\xspace}

\begin{document}
\LARGE{Neutrino experiments in  Physics Department of Rome ' Sapienza'University}
\vskip 1cm
\vskip 1.cm
\par
\Large{Ubaldo Dore \\
Dipartimento di Fisica, Universit\`a di Roma ``Sapienza",\\
and I.N.F.N., Sezione di Roma, P. A. Moro 2, Roma, Italy\\}
\newpage

\vskip 1cm
\par ABSTRACT 
\par
This paper will describe the history of experimental neutrino 
physics in the 
 Physics Department of the Rome  'Sapienza' University.    
\newpage
\tableofcontents
\input{Intro1}

\input{history}

\input{newresu}

\input{conclu}
\input{ack}
\bibliographystyle{unsrt}
\bibliography{newneu}
\end{document}

%% file: Intro1.tex
\newpage
\section{Introduction}
This article will describe the development of neutrino physics in 
the Physics Department of the  'Sapienza' Rome University.
\par The  "Istituto Nazionale di 
Fisica Nucleare" (INFN) supported  particle physics activity since
its  foundation. 
Members 
of the Department and of the INFN Section 
proposed, participated and led  
 the design, construction and data analysis of neutrino experiments.
 \par Particle Physics at he Sapiehnzabegan  when Fermi and his group 
moved in 
the 
newly built Department in 
1936. After the departure of Fermi, due to the  racial laws, 
particle and nuclear physics continued 
during  the difficult years of the war. In these years  Marcello 
Conversi, Ettore Pancini and Oreste  Piccioni performed a fundamental 
experiment on the nature of the hard component of cosmic rays discovering 
the heavy electron, the muon.The muon was the heavier replica of the 
electron.
There were experiments to study  cosmic rays, in the Italian 
Alps, with balloons flights and in laboratories.
In  1975 Marcello Conversi proposed  studying  charm particles 
produced by neutrino beams. In the sixties and seventies bubble chamber 
experiments were performed with the  participation of Rome experimental 
groups. 
\par From the seventies Neutrino experiments were performed 
with electronic detectors at CERN, Japan and USA.  
In addition to accelerator experiments in Rome physics department we have 
experiment on the
Majorana or Dirac nature of neutrinos and experiments measuring neutrinos  
from astrophysical sources.  
\par We will describe the participation of the 
'Sapienza' University and INFN in these experiments.
\par The  theorist contribution to neutrino physics is given in next 
section
\subsection {Theoretical contributions to neutrino physics }  
\par Important theoretical contributions have been made in the past in 
the Department ( before Istituto)
 and are still going on. 
Main fields of research in   
particle physics are : weak interactions, elecroweak theory and 
atmospheric and cosmic radiation.
\par For what concerns neutrino physics we give numbers 
 (approximate) of published 
 papers and arxiv preprint hep-ph 
in authors alphabetic order a also mentioning a few  selected   papers. 
\par Guido Altarelli (4 papers) Must  Heavty Leptons have their own 
neutrino (Phys. Lett. B67 463, 1977. Charmed quarks and asymptotic freedom 
in
neutrino scattering  (phys LettB 48 435 1974)
 Altarelli has worked in Rome until 1976, he has then continued his 
activity in 
CERN  and Roma3 University.  
 \par Nicola Cabibbo (9 papers): CP violation in neutrino interactions 
(XII Neutrino Telescope Workshop, 1978). Time reversal 
 violation in neutrino interactions (Phys. Lett. B72 333 1978).
Neutrino processes in a compound model for the nucleon (Phys. Lett. B48 
435 
1974). ( Paper in collaboration With L. Maiani and G. Altarelli)
\par Daniele Fargion. (19 papers)
Ultra high energy neutrino scattering (Astrophysics J. 517 725 1999).
Discovering ultra high energy neutrinos (Astrophysics J. 570 908 2002)
\par Paolo Lipari (21  papers):
 The neutrino cross section of upward going muons( Phys. rev. 
Lett.74,4384,1995). Comparison of $\numu \rightarrow \nutau$ and $\numu 
\rightarrow \nu_{s}$ (Phys. Rev. D58 073005 1998)
\par Maurizio Lusignoli(24 papers): Fresh look at neutrino oscillations  
(Nuclear Physics B168 1980).Neutrino decay and atmospheric neutrinos
(Phys. lett. B462 109 1999)
\par Luciano Maiani (11 papers)
Theory of interactions of neutrino with matter. (Neutrino physics, editor 
K. Winter, pg 278 1981) .Elecromagnetic corrections to neutrino processes
(Nucl. Phys. B186 1981)
\par Important contributions were also made by  Guido Martinelli, Barbara 
Mele  and Silvano  Petrarca.

%% file: history.tex
\section{History}

\label{charms}
The Department of Physics (initially Istituto) was built in 1935 and the 
Fermi group moved in from the old Institute of Via Panisperna.
The group was composed of 
E.Fermi, E.Amaldi, B.Pontecorvo, F.Rasetti and E.Segre.\par The impact of 
Pontecorvo on neutrino physics, with  his activity  described in 
\cite{dore}, has been widely  
recognized in the scientific world. For example V.Telegdi once said
\par{\sl
Almost all important ideas in neutrino physics are due to Pontecorvo} 
\par In the thirties, two fundamental contributions to particle physics 
were 
made: 
 Fermi developed the theory of beta decay \cite{beta} and in 1934 
Fermi and his group realized the fission of uranium irradiated with 
neutrons\cite{fermi1,fermi2}.   
Particle physics  started with the Conversi, Pancini, Piccioni experiment
that showed that the cosmic rays particles were not strongly 
interacting particles. They were weakly interacting particles 
called 'muons'.
The importance of the experiment was such that it was recognized 
as the origin of particle physics 
 by Luis Alvarez in his 1978 Nobel lecture. 
Neutrino physics started in Rome in 1975 when Marcello Conversi proposed 
an 
experiment for the detection of charmed particles in a neutrino beam.   
 \cite{conversi1} 
Experiments  with neutrino beams were performed at CERN and in the  United 
States in the '70s and '80 mainly using bubble chambers.
These were Gargamelle (1970-1978) and BEBC installed at the beginning of 
the 1970  and dismounted in 1985. In the late seventies   
big electronic detectors devoted to neutrino physics started :
\par  CDHS(1977-1983)and  CHARM(1978-1984) were devoted to the study of 
deep inelastic 
neutrino scattering (charged and neutral currents)
\par CHARM2 (1986-1990) was built to study neutrino electron elastic 
scattering.
\par The main purpose of Nomad(1994-1997) and Chorus(1992-1997)  a 
hybrid 
experiment,
was to 
look for   neutrino oscillations \ref{bruno}.
\par Rome groups participated in the 
 CHARM, CHARM2 and CHORUS experiments. 
\par Long
baseline experiments, are  experiments in which a detector is placed  at 
several
hundred km from the source of neutrinos. The large distance is needed  to 
access the small $\Delta m^{2}$ 
region of oscillations observed
in atmospheric neutrino experiments. Rome has participated in  several 
experiments,
 K2K(1999-2004) designed to confirm results of atmospheric neutrinos  at 
KEK in 
Japan. Rome is still participating in  
\par Opera(2008-):Observation of neutrino tau in a beam of neutrini mu 
CERN-Gran Sasso 
\par T2K(2009-)
 at Jpark Japan:  detection of  $\numu$ to $\nue$ oscillations
Finally  Rome is involved in the  CUORE(2008-) experiment searching  for 
double 
beta decay and in 
astrophysical neutrino measurement in large underwater detectors, 
NEMO(1998-)

%% file: newresu.tex
\section{results and experiments}
\par This section will contain results of the experiments
described in the previous section, in particular 
\begin {itemize}
\item {charm physics results}
\item{oscillations :theory,chorus,K2K,T2K and Opera}
\item {Large detector results:  charm,charm2}
\item {Fermilab :Sciboone results}
\item {Cuore  :Search for double beta decay}
\item  {Nemo :Detection of extraterrestrial neutrinos}
 \end {itemize}
\par 
In the end we will recall a proposal, largely due to Rome,
to check the oscillation  
result obtained by the 
LSND \

\subsection{Study of charm physics}
\label{charms}
The Discovery in 1974 of J/$\Psi$ particle, a  charm-anti charm state and 
then of particles containing a charm quark, were 
considered as 
proof of the existence of a new particle the Charm (C).
The direct observation of their flight path was a challenge because 
the charmed  particles had to decay 
weakly with lifetimes of 10$^{-14} s,
10^{-15}$ with decay path $\lambda$ length of few $\mu$m ( $\lambda =  
c\tau\gamma$). 
\par  The final proof 
of the existence of these particle had to be  the measurement
of their path. For this reason, Marcello Conversi  proposed to observe 
the short life charm particles produced in neutrino interactions   
in  experiments
combining  nuclear emulsions, bubble chambers and counter
techniques.\cite{conversi1}. In fact, being a weak process, neutrinos 
interactions violate charm conservation and the charm particles  particles 
can be produced singly  in the reaction
$\numu + N \rightarrow \mu +1 charmed +other hadrons$
Experiments were made at CERN And Fermilab. In Europe a large emulsion  
collaboration was set up. Rome did participate under the guidance of 
prof. 
Giustina Baroni. Lifetimes of Charm particles were measured for mesons 
and hadrons
\subsection {Neutrino Oscillations}
\par In this section after an introduction to neutrino oscillations we 
will 
consider the following experiments: Chorus 
Opera ,K2K,
T2K to  which Rome contributed.
\par The value  of SuperKamiokande  $\Delta m^{2}$ for  $\numu$
oscillations requires values of
  L/E km/Gev of the order of
10$^{3} $km/Gev to  detect oscillations, recall formula (1) of
section \ref{bruno}. For this reason Opera,K2k,Te2k are long baseline 
experimens.
\subsubsection {  Neutrino oscillations}  
\label{bruno}
 Neutrino oscillations, first 
predicted by 
Bruno Pontecorvo and then observed experimentally, are a 
quantum  mechanical process in which a neutrino created with a 
certain flavor can be observed later with a different flavor.
The flavor eigenstates are not the mass eigenstates. The mixing matrix
can be expressed in terms of   three  mixing angles and a phase factor.
\par The present  experimental values for the mixing angles
are\cite{datap}
\par $sin^{2}2\vartheta_{12}=0.857\pm 0.024$
\par $sin^{2}2\vartheta_{23}> 0.95$
\par $sin^{2}2\vartheta_{13}=0.098\pm 0.0.012$.
\par The latter  result  has been obtained  in the 
last 
two years using data from reactor 
experiment :Doublechooz \cite{double},Daya Bay \cite{daya} ,Reno 
\cite{reno} , 
accelerator:T2K  \cite{t2knue}. 
\par Before there  was  
only an upper limit CHOOZ
\par $sin^{2}2\vartheta_{13}<0.13$ \cite{apol} 
\par Oscillations are characterized by the mass difference for different 

species $\Delta m_{il}^{2} =m_{i}^{2}-m_{l}^{2}$
\par For three neutrinos t only two independent mass differences are 
possible and measured.  The smallest 
$\Delta m^{2}_{12}= (7.9.\pm 0.06).10^{-5} eV^{2}$  is 
measured in 
 solar and  reactor
neutrino experiments and the largest $\Delta m^{2}_{23}=(2.4\pm 0.21). 
10^{-3} eV^{2}$
 \par is measured  in  atmospheric and accelerators
  neutrinos
experiments.
We recall that the oscillation probability fo only two 
 neutrinos
can be written as $$P=sin^{2}2\vartheta sin^{2}(1.27 \Delta m^{2}L/E)  
~~~(1)$$
where $\vartheta$ is the mixing angle   and $\Delta m^{2}$ is the  
mass squared 
difference of the two neutrinos, L is the distance between source and 
detector and E the energy of neutrinos.
In the three neutrino mixing scheme  the probability 
of $\numu$ to $\nue$ oscillations can be written in first approximations 
$$P=sin^{2} 2\vartheta_{13}sin^{2}( \vartheta_{23})sin^{2}(1.27
 \Delta m^{2}_{23}L/E)~~~~~~    (2)$$
where
$\vartheta_{23}$ and $\Delta m_{23}^{2}$ refer to the mixing of the \numu 
to 
\nutau 
as seen in atmospheric and accelerators disappearance experiments
while $\vartheta_{13}$ is observed in  \nueb reactor experiments 
disappearance 
experiments as well in the recently observed   \numu to \nue oscillations
\par The complete formula of the oscillation probability (first 
approximation in (2)) that can be found, 
for example in \cite{bart}, it contains 
 the phase factor of the mixing matrix $\delta$ and 
terms that are sensitive to the mass hierarchy and matter effects.
\subsubsection{CHORUS}
The search for neutrino oscillation made in CHARM and CHARM2 did not 
produce any positive result.
\par The collaboration proposed an  experiment  in 1990 to 
look with high sensitivity 
for \numu to \nutau oscillation.The experiment was approved in 1991  
and the 
construction of the detector  was completed in 1994.
Data taking started in the same year  and was completed in 1998. The 
description of 
the apparatus  can be found \cite{chorusdet}. The detector was a hybrid 
design 
with a large emulsion target (770 kg). Planes of high resolution fiber 
trackers to predict impact point of track on the emulsions were followed 
downstream   by an 
electromagnetic  calorimeter and a muon spectrometer. The calorimeter 
\cite{calor} was built by 
italian groups (Ferrara,Napoli,Roma)
\par At that time there were indication for 
oscillation in the  deficit   of neutrino emitted by the sun , while there was no 
evidence  for \numu oscillation. A
motivation of the experiment can be found in ref.\cite{harari},
where high values 
of   $\Delta M^{2}$ and low mixing  for \numu to \nutau oscillation
were 
predicted. This 
mass  region was accessible in an high sensitivity   experiment at  CERN SPS neutrino Beam 
 where sensitivity of $10^{-4}$ in the oscillation probability 
 in the high mass region 
  could be obtained    in few years of running.
\par The final result of the experiment was published in ref \cite{final}
giving a limit of $sin^{2}\vartheta _{\numu-\nutau}$~ of~ $4.4.10^{-3}$ 
 for large$\Delta m^{2}$. 
\par Because of the high granularity of emulsions and the development of 
fast automatic measuring systems,  a large amount of charm events (2059)
were found and the decays  
classified according to their topology. About 15 papers were published from 
1998 and 2011. They included  :
Cross sections of charm production in anti neutrino
 interactions\cite{anti},
properties of the D$^{0}$ meson \cite{frag} , associated charm production 
\cite{assoc2} , measurement of muonic branching ratios of charmed 
hadrons \cite{topo}

\subsubsection{OPERA}
\par Neutrino \numu oscillations have been first observed in the 
disappearance mode of atmospheric neutrinos
experiments. The overall picture of atmospheric, solar, accelerators and 
reactor experiments is consistent with the fact that the disappearance is 
due to  \numu to \nutau
oscillation, but no direct proof has been obtained. OPERA is an 
experiment to detect 
\nutau from oscillations in a pure \numu beam using high resolution 
nuclear emulsions that 
allow a measurement of the $\tau$ decay. 
 The \numu beam from CERN is sent to  LNGS
at a distance of 730 km where Opera is located.
The average neutrino beam energy is 17 GeV which does not correspond to 
the 
maximum of oscillations but rather to the optimization of  of the 
number of observable $\tau$
taking into account
the threshold for $ \tau$ production. 
The experiment detector consist of 150,000 bricks of ECC ( emulsions +lead 
foils 
sandwiches) 
for a 
total 
mass of 1300 tons. The experiment is complemented with trackers to  
reconstruct charged tracks and locate the interaction vertex
 and followed by a muon spectrometer to measure muon  momentum.
The Rome group led by  G.Rosa participates in the analysis of the  
emulsions. 
\par The experiment started data taking  in 2008 and is still running. At 
the 
moment two neutrino tau interactions have been detected \cite{nakamura}.
with a  background of 0.2 event   
 \subsubsection{K2K}

\par In 1998 evidence was presented  for \numu oscillation in atmospheric 
neutrinos \cite{Fuku}. In 1999 a long baseline accelerator experiment was 
started to confirm this  oscillations at an accelerator neutrino beam 
this phenomenon. Neutrino of about 1 GeV were generated using the KEK 
accelerator and 
observed in the SuperKamiokande large water Cherenkov  detector at a 
distance of 275  km as the far detector. The 
experiment had   a two 
detector layout described in  \cite{Han}. The 
near  detector was composed of  a large water Cherenkov detector, 
scintillation fiber trackers ,a fine grained scintillator detector, an 
electromagnetic calorimeter and a muon ranger.  
The electromagnetic calorimeter for the Run II of the experiment was built 
by reusing modules of the  
CHORUS calorimeter.The installation of the calorimeter, calibration and 
running of the experiment was responsibility of the Rome group (led by L. 
Ludovici) which joined the experiment in 2002.
\par At the end of data taking in 2005  112 events were 
observed against an expectation of 
158 in the absence of oscillations. The  \numu disappearance interpreted 
in  
terms of  oscillation gave a 
value of $\Delta m^{2}$ in agreement with the values indicated by 
 atmospheric neutrino  experiments.  \cite{k2k}

\par Moreover several results   have been published on neutrino cross 
sections 
using data 
of the close 
detector, we 
recall here the results on charged pion production \cite{Rodriguez}
coherent pion production \cite{Hasegawa}
the analysis of $\pi^{0}$ production \cite{mariani}

  \subsubsection{T2K} 
\par Until few years ago the value of $\vartheta_{13}$ was unknown and
 only upper limits were available  \cite{apol} The knowledge of  
value of this quantity is  important because a not too small value opens 
 the possibility of measuring the CP violating   phase 
$\delta$ appearing in the 
mixing matrix. Many experiments are  searching for this value,     
both reactor experiment to observe the disappearance of \nueb and 
accelerator experiment 
looking for \nue appearance in a \numu  beam 
oscillations that are possible in a 3 flavor mixing scenario see 
formula (2) in section \ref{bruno}.
T2K is a search for \nue appearance in a neutrino  beam produced at JPARC 
 numu appearance experiment.  
As in K2k, the far detector is SuperKamikande which sees an off-axis 
neutrino beam with 0.6 average neutrino energy at a baseline of 296 km
tuned to the first maximum of atmospheric neutrino oscillations.
The near 
detector located at a distance of 280 meters from the target   
consists of 
three large 
volume time projection chambers interleaved with two fine grained 
scintillator tracking detectors and a scintillating detector dedicated to
$\pi^{0}$ detection. All of the system is embedded the CERN UA1 magnet. 
The
idea of magnetizing the close detector by refurbishing the CERN UA1  
magnet 
was  suggested by PF Loverre (Rome).
The Rome group (leader Lucio Ludovici) is now working in the analysis
\par  The first result based on 6 candidates was published in 2011 
\cite{t2knue} and the present result  ,based on 11 \nue 
events, was  presented at the 2012 
neutrino conference
\cite{nakaya}
$$ sin^{2}2\vartheta_{13}=0.104+0.06-0.045 $$ 
\par The present result is based on a value of $\delta$= 0 and of normal 
hierarchy.
\subsection{results obtained with general purpose detectors}
\subsubsection{CHARM}
 The CHARM detector was    primarily
designed 
to investigate neutral-current interactions of high-energy neutrinos.
. The detector, described in \cite{Diddens} consisted of an ionization 
calorimeter 
 target
to determine the energy and the direction of the shower produced by the   
neutrino scattering, and of a magnetic muon spectrometer.
The detector was  run in the SPS neutrino beam from 1978 to 1982.
The spokesman of the CHARM collaboration  (acronym for CERN, 
Hamburg,Amsterdam,Rome,Moscow )
was  Klaus Winter and the Rome group was led by Bruno Borgia.                                                
The Rome group contributed to the construction of the detector, to the 
running of the experiment and to the analysis of the data.
Results were obtained in various fields and  many papers were published.
We recall the measurement of the electroweak mixing angle, a fundamental 
parameter of the electroweak theory. In  neutrino 
semi leptonic interactions the collaboration 
           measured from the ratio of neutral over charged current 
interactiona     
$$ R=\frac{\sigma_{NC}}{\sigma_{CC}}=
 1/2-sen^{2}\vartheta_{w}+5/9 +sen^{4}\vartheta_{w}(1+r)$$
where r is ratio of neutrino and anti neutrino cross sections
a value of $$sin^{2}(\vartheta_{w}) =.236+0.012(m_c-1.5)\pm.005(exp)\pm 
0.003 (theor)$$
\cite{allaby}. It must be noted that the result has been parametrized 
in term of $m_c$ ,the charm quark  mass.  
 \par The collaboration  also measured  the  value of the mixing angle 
from the ratio of the cross sections of neutrino and anti neutrino 
electron scattering
\cite{dorem}
$$R\frac{\sigma(\nu_{\mu} 
e)}{\sigma(\nubar_{\mu} e)}=3\frac{(1-4sin^{2}(\vartheta_{w})
+16/3 sin^{2}(\vartheta_{w})}{
(1-4sin^{2}(\vartheta_{w})+16/sin^{2}(\vartheta_{w})}$$ obtaining  a value 
of $$sin^{2}(\vartheta_{w}) =.211\pm.037(stat)$$ .   
A total of 83 events in the neutrino  and 112 in the anti neutrino  
beam were observed.
\par Results were also obtained in the beam dump layout \cite{prompt}.
In this layout  the proton beam was stopped in a heavy target (copper) 
to study  prompt decays. Prompt decays were expected to be mostly to  charm decays
as longer living  
hadrons would have interacted  before decaying. 
The result were consistent with the assumption that the 
prompt neutrinos did come from charm decays.
\par A two detector layout  was made in the 
PS neutrino beam \cite{pstwo} searching for neutrino oscillations.
A search was also made in the SPS beam \cite{bergsma}. Both searches were 
negative and upper limits were set as results.  
\subsubsection{CHARM2}
CHARM2 was  designed especially to study neutrino electron scattering. 
\par The target material was glass and it was 
was a massive (692 t)  high granularity low Z detector.	It is described in 
\cite{dewinter}. Its spokesman was Klaus Winter
and the Rome group was led by U. Dore.
\par The italian  group contributed to the construction of the limited        
streamer counters. They were built in Italy under the
supervision of the Rome group.
\par  The study of neutrino electron
scattering  allowed  a precise determination of the electroweak mixing
from the ratio of the cross sections of neutrino and anti neutrino electron 
scattering.
The ratio of cross sections was determined by counting the number
of electron events in the neutrino (3886) and anti neutrino (4996) beams.  
\cite{tetaw}
A value of $$sin^{2}(\vartheta_{w}) =.2324\pm.0058(stat) \pm
0.0059(syst)$$ was determined.
\par Results were also obtained on the study of dimuons,  events with 
a first muon  produced in a charged current interaction and a second one 
coming from the muonic decay of a charmed  particle 
\cite{dimuon}, on the  determination of the cross section of the 
inverse muon decay \cite{inverse} and on the search of neutrino 
of  $\numu$  to $\nutau$ oscillation \cite{gruwe}.
Because the short lifetime of the $\tau$ produced by the charged 
current interactions of oscillated $\nutau$  the granularity of the 
detector did not allow a direct observation,  topological 
arguments were  used; reaching the upper limit  of $3.4.10^{-3}$ for 
$sen^{2}\vartheta$ 
for  $\Delta M^{2}$ =25 e$V^{2}$ A different approach has been proposed, 
oscillations 
reduce the number of charged currents. For large $\Delta m^{2}$ 
an upper limit 
$sen^{2}\vartheta \leq  3. 10^{-3}$ was set
\subsection {SCIBOONE}
\par The experiment was  dedicated to the measurement of neutrino an anti 
neutrino cross section at around 1 GeV energy. The experiment was  
done  at the booster neutrino beam of the   Fermi National Laboratory 
(FERMILAB). The data taking started in 2007 and finished in 
2008. The Italian group leader was L. Ludovici.
The experiment consisted 
of three subsystem: a fine grain tracker (Scibar), an 
electromagnetic 
calorimeter and a muon range detector.
Scibar and  the electromagnetic calorimeter constructed for the K2K 
experiment were  
transported to  Fermilab and reused.
\par Results have been obtained on 
neutrino and antineutrino interactions\cite{NaKaji},
neutral $\pi^{0}$ \cite{pi0}, coherent pion production in 
charged \cite{Hiraide} and neutral \cite{Kurimoto} current
production  ,$K^{+}$ \cite{kap}
production. Results on $\numu$ and $\numub$ disappearance were  
obtained in a   
combined experiment Sciboone +Miniboone  \cite{Mahn}.
\subsection {CUORE}
\label{cuoricino}
\par The assignment of the   neutrino nature,  whether they are
Dirac or Maiorana particles,
is still an open problem. The CUORE experiment, search
of the neutrinoless beta decay will tackle this problem . This process is
in fact  permitted for Maiorana neutrinos and forbidden for Dirac ones.
The experiment situated in the  Laborarori del  Gran Sasso (LNGS)  cavern
is
a large cryogenic detector
built of $T_{e}O_{2}$ crystals cooled at 10 mK. Results obtained in a test
apparatus Cuoricino have been published in 2006 \cite{cuo}.
\subsection{NEMO}
\par The search for  extra galactic high energy neutrino sources is the
aim
of the NEMO underwater Cherenkov  neutrino detector \cite{amore1}.The
construction of
a km$^3$ detector in the northern emisphere  is  a  goals of astro
particle  physics. The Rome group (leader a. Capone) is  participating to 
the experiment.
As opposed  to photons and charged particles neutrinos interact only 
weakly  
and so they proceed in 
straight lines thus giving  information on galactic and
extragalactic sources. The Nemo detector, situated at a depth of 3500 m
will be located in the
mediteranean sea, at capo Passero near Sicily at a depth of 3500 m. The 
detector will have of 
the order of 100 vertical towers equipped with photomultipliers.

 \subsection{An experiment to check the LSND results}
In 1996 the LSND collaboration presented results showing some  evidence  
of
\numub to \nueb
oscillations at a $\Delta m^{2}$ of the order of eV$^2$.  Their final 
result can 
be found in \cite{LSND}. Because there were
already the indications of a two values of  $\Delta m^{2}$ 
,see section \ref{bruno}
from solar  neutrinos and  from atmospheric neutrino 
experiments, with  a three neutrino model there 
was no 
room for a third value 
 of $\Delta m^{2}$. This  result, if confirmed, would  
require a fourth neutrino, which had to be sterile because of LEP 
limit on the light active neutrinos.
 Rome had a leading role 
in setting  up  a large collaboration \cite{arme}  aiming 
 to firmly confirm or refuse  the LSND results.
\par A two detector experiment  at CERN was designed, with a  neutrino 
bean  
sent 
from the PS to  a close detector at 130 m  and a far one at 880 m.
\par The two  detectors consisted of scintillation bars interleaved with 
iron sheets . It was a fine grain high sensitivity apparatus suitable to 
detect 
\nue. The proposal was  rejected  by the SPSC 
committee.
It must be noted that the proposal was valuable, in fact
 after 15 years a similar proposal has 
been made from the Carlo Rubbia group \cite{anton}
This rejection of the proposal   
was noxious  for  the neutrino group in Rome,
 a large majority
of persons involved in the proposal left neutrino physics.

%% file: conclu.tex
\section{Conclusions}
This paper describes the evolution of experimental neutrino Physics in 
the Rome Sapienza Physics Department.
After 40 years  the activity is still continuing with Cuore, Opera, Nemo 
and T2K experiments. 

%% file: ack.tex
\section{Ackwledgement}

I warmly thank Lucio Ludovici for encouragement, criticism, discussion 
and final reading. I gratefully aknowledge the critical reading of Bruno
Borgia, Maurizio Lusignoli,  Peter Nemethy and Graham Hymus  